# Quantum dialogue based on quantum encryption with single photons in both polarization and spatial-mode degrees of freedom

Tian-Yu Ye*, Mao-Jie Geng, Tian-Jie Xu, Ying Chen

College of Information & Electronic Engineering, Zhejiang Gongshang University, Hangzhou 310018, P.R.China

**Abstract**

In this paper, a novel information leakage resistant quantum dialogue (QD) protocol with single photons in both polarization and spatial-mode degrees of freedom is proposed, which utilizes quantum encryption technology to overcome the information leakage problem. In the proposed QD protocol, during the transmission process, the single photons in both polarization and spatial-mode degrees of freedom used for encoding two communicants' private classical bits are protected by both quantum encryption technology and decoy photon technology. For avoiding the information leakage problem, the initial states of the single photons in both polarization and spatial-mode degrees of freedom used for encoding two communicants' private classical bits are shared between two communicants through quantum key encryption and decryption. The information-theoretical efficiency of the proposed QD protocol is as high as 40%.

## 1 Introduction

It is well known that since the birth of quantum cryptography [1], it has gained many interesting branches, such as quantum key distribution (QKD) [1-4], quantum private comparison (QPC) [5-9], quantum secure direct communication (QSDC) [10-14], etc. It is natural that different branches have different functions. As for QKD, its goal is to create a random raw key between two remote communicants by virtue of the transmission of quantum signals. With respect to QPC, it aims to judge whether the private inputs from two users are equal or not. With regard to QSDC, it concentrates on directly transmitting private messages from one user to the other user without creating a random raw key to encrypt it in advance.

In fact, many earlist QSDC protocols just accomplished the unidirectional transmission of a private message from the sender to the receiver, which maight be useless in the mutual dialogue circumstance. As a result, how to acheive the bidirectional transmission of private messages between two communicants became urgent to solve. Fortunately, in the year of 2004, Zhang et al. [15-16] and Nguyen [17] independently suggested the completely novel concept named quantum dialogue (QD) to solve this problem. Hereafter, many scholars threw their enthusiasms into the research of QD. Consequently, lots of QD protocols [18-24] have been constructed. Later, a sepecial security loophole named information leakage was independently discovered by Tan and Cai [25] and Gao et al.[26-27], which means that an outside Eve can easily deduce out partial information about the private messages of communicants from the public announcement without launching any active attacks. Hence, how to effectively solve the information leakage problem became of utmost importance in the realm of QD. In the year of 2009, Shi et al. [28] utilized the method of directly transmitting shared private Bell states from one communicant to the other communicant to overcome the information leakage problem. In the year of 2010, Shi et al. constructed a QD protocol using the direct transmission of shared private single photons to avoid the information leakage problem [29] and a QD protocol without information leakage based on the correlation extractability of Bell state and the auxiliary single particle [30]; Gao [31] utilized the measurement correlation from the entanglement swapping between two Bell states to resist the information leakage problem. In the year of 2013, we proposed a large payload QD protocol without information leakage through the entanglement swapping between any two GHZ states and the auxiliary GHZ state [32]. In the year of 2014, we designed a quantum encrypting QD protocol without information leakage [33]. In the year of 2015, we used auxiliary quantum operation to solve the information leakage problem [34].

Recently, many scholars adopted single photons in both polarization and spatial-mode degrees of freedom to design different kinds of quantum cryptography protocols [35-38]. The reason lies in that in a quantum cryptography protocol, if single photons in both polarization and spatial-mode degrees of freedom are used as quantum resource instead of single photons with one degree of freedom, its quantum communication capacity is usually greatly enlarged.

Based on the above analysis, in this paper, we put forward a novel QD protocol without information leakage by adopting single photons in both polarization and spatial-mode degrees of freedom as quantum resource, which utilizes quantum encryption technology to overcome the information leakage problem. In the proposed QD protocol, one single photon in both polarization and spatial-mode degrees of freedom can totally carry four classical bits, which means that it has a large quantum communication capacity.

The remaining parts of this paper are arranged as follows: Sect.2 introduces preliminary knowledge; Sect.3 describes the proposed QD protocol; Sect.4 conducts the security analysis; and finally, Sect.5 gives discussions and conclusions.

## 2 Preliminary knowledge

---

*Corresponding author:
E-mail：happyyty@aliyun.com(T.Y.Ye)

A single-photon state in both polarization and spatial-mode degrees of freedom can be described as [35]

$$|\phi\rangle = |\phi\rangle_P \otimes |\phi\rangle_S. \quad (1)$$

Here, $|\phi\rangle_P$ and $|\phi\rangle_S$ are the single-photon states in the polarization and the spatial-mode degrees of freedom, respectively. $Z_P = \{|H\rangle, |V\rangle\}$ and $X_P = \{|R\rangle, |A\rangle\}$ are two nonorthogonal measuring bases in the polarization degree of freedom, while $Z_S = \{|b_1\rangle, |b_2\rangle\}$ and $X_S = \{|s\rangle, |a\rangle\}$ are two nonorthogonal measuring bases in the spatial-mode degree of freedom. Here,

$$|R\rangle = \frac{1}{\sqrt{2}}(|H\rangle + |V\rangle),\ |A\rangle = \frac{1}{\sqrt{2}}(|H\rangle - |V\rangle),\ |s\rangle = \frac{1}{\sqrt{2}}(|b_1\rangle + |b_2\rangle),\ |a\rangle = \frac{1}{\sqrt{2}}(|b_1\rangle - |b_2\rangle). \quad (2)$$

$|H\rangle$ and $|V\rangle$ are the horizontal and the vertical polarizations of photons, respectively, while $|b_1\rangle$ and $|b_2\rangle$ are the upper and the lower spatial modes of photons, respectively.

Define two unitary operations in the polarization degree of freedom as

$$I_P = |H\rangle\langle H| + |V\rangle\langle V|, U_P = |V\rangle\langle H| - |H\rangle\langle V|. \quad (3)$$

It turns out that

$$I_P|H\rangle = |H\rangle, I_P|V\rangle = |V\rangle, I_P|R\rangle = |R\rangle, I_P|A\rangle = |A\rangle, \quad (4)$$

$$U_P|H\rangle = |V\rangle, U_P|V\rangle = -|H\rangle, U_P|R\rangle = -|A\rangle, U_P|A\rangle = |R\rangle. \quad (5)$$

Apparently, $U_P$ only flips the photon inside its base. Likewise, define two interesting unitary operations in the spatial-mode degree of freedom as

$$I_S = |b_1\rangle\langle b_1| + |b_2\rangle\langle b_2|, U_S = |b_2\rangle\langle b_1| - |b_1\rangle\langle b_2|. \quad (6)$$

It turns out that

$$I_S|b_1\rangle = |b_1\rangle, I_S|b_2\rangle = |b_2\rangle, I_S|s\rangle = |s\rangle, I_S|a\rangle = |a\rangle, \quad (7)$$

$$U_S|b_1\rangle = |b_2\rangle, U_S|b_2\rangle = -|b_1\rangle, U_S|s\rangle = -|a\rangle, U_S|a\rangle = |s\rangle. \quad (8)$$

Apparently, $U_S$ only flips the photon inside its base.

## 3 Description of the proposed QD protocol

There are two communicants, Alice and Bob, who want to exchange their private classical bits. Suppose that Alice's $2N$ private classical bits are

$$\{(t_1, j_1), (t_2, j_2), \cdots, (t_i, j_i), \cdots, (t_N, j_N)\}, \quad (9)$$

while Bob's $2N$ private classical bits are

$$\{(k_1, l_1), (k_2, l_2), \cdots, (k_i, l_i), \cdots, (k_N, l_N)\}, \quad (10)$$

where $t_i, j_i, k_i, l_i \in \{0,1\}, i \in \{1,2,\cdots,N\}$. They agree on beforehand that each composite unitary operation corresponds to two classical bits in the way as

$$C_{00} = I_P \otimes I_S \to 00, C_{01} = I_P \otimes U_S \to 01, C_{10} = U_P \otimes I_S \to 10, C_{11} = U_P \otimes U_S \to 11. \quad (11)$$

The proposed QD protocol can be depicted as follows.

**Step 1: Quantum key generation and distribution.** Alice prepares $N$ four-particle entangled states $\{A_1^1 A_2^1 B_1^1 B_2^1, A_1^2 A_2^2 B_1^2 B_2^2, \cdots, A_1^N A_2^N B_1^N B_2^N\}$ all in the state of

$$|\psi\rangle = \frac{1}{2}(|H\rangle|H\rangle|H\rangle|H\rangle + |H\rangle|V\rangle|H\rangle|V\rangle + |V\rangle|H\rangle|V\rangle|H\rangle + |V\rangle|V\rangle|V\rangle|V\rangle), \quad (12)$$

where the superscripts $1, 2, \cdots, N$ denote the order of these four-particle entangled states. Alice splits these four-particle entangled states into four particle sequences $S_{A_1}, S_{A_2}, S_{B_1}, S_{B_2}$, respectively, i.e., $S_{A_1} = \{A_1^1, A_1^2, \cdots, A_1^N\}$, $S_{A_2} = \{A_2^1, A_2^2, \cdots, A_2^N\}$, $S_{B_1} = \{B_1^1, B_1^2, \cdots, B_1^N\}$ and $S_{B_2} = \{B_2^1, B_2^2, \cdots, B_2^N\}$. For the sake of security check, Alice generates two sets of decoy photons, each of whose particles is randomly in one of the four states $\{|H\rangle, |V\rangle, |+\rangle, |-\rangle\}$, where $|+\rangle = \frac{1}{\sqrt{2}}(|H\rangle + |V\rangle)$ and $|-\rangle = \frac{1}{\sqrt{2}}(|H\rangle - |V\rangle)$. Then, Alice randomly inserts one set of decoy photons into $S_{B_1}$ to form particle sequence $S_{B_1}'$ and sends it to Bob with the block transmission method [10]. Likewise, Alice randomly inserts the other set of decoy photons into $S_{B_2}$ to form particle sequence $S'_{B_2}$ and sends it to Bob with the block transmission method.

After Bob confirms the receipt of $S'_{B_1}$ ($S'_{B_2}$), Alice informs Bob of the positions and the preparation bases of the decoy photons in it. Then, Bob measures the corresponding decoy photons with the correct measuring bases and tells Alice his measurement results. Alice calculates the error rate by comparing Bob's measurement results with the initial states of decoy photons. If the transmission of $S'_{B_1}$ ($S'_{B_2}$) is secure, they will continue the next step; otherwise, they will halt the communication.

**Step 2: Alice's encryption.** Alice generates a sequence of $N$ single-photon states in both polarization and spatial-mode degrees of freedom, i.e., $S_T = \{|\phi\rangle_1, |\phi\rangle_2, \cdots, |\phi\rangle_N\}$, where the subscripts $1, 2, \cdots, N$ denote the order of these single-photon states in $S_T$, $|\phi\rangle = |\phi\rangle_P \otimes |\phi\rangle_S$, $|\phi\rangle_P \in \{|H\rangle, |V\rangle\}$ and $|\phi\rangle_S \in \{|b_1\rangle, |b_2\rangle\}$. Alice measures partciles $A_1^i$ and $A_2^i$ with $Z_P$ basis, respectively, where $i = 1, 2, \cdots, N$. In order to encrypt the single-photon states in $S_T$, Alice performs the composite unitary operations on them according to the following rules: if the measurment results of partciles $A_1^i$ and $A_2^i$ are $|H\rangle$ and $|H\rangle$, respectively, Alice performs the composite unitary operation $C_{00}$ on $|\phi\rangle_i$; if the measurment results of partciles $A_1^i$ and $A_2^i$ are $|H\rangle$ and $|V\rangle$, respectively, Alice performs the composite unitary operation $C_{01}$ on $|\phi\rangle_i$; if the measurment results of partciles $A_1^i$ and $A_2^i$ are $|V\rangle$ and $|H\rangle$, respectively, Alice performs the composite unitary operation $C_{10}$ on $|\phi\rangle_i$; and if the measurment results of partciles $A_1^i$ and $A_2^i$ are $|V\rangle$ and $|V\rangle$, respectively, Alice performs the composite unitary operation $C_{11}$ on $|\phi\rangle_i$. Then, Alice prepares enough decoy single photons in both polarization and spatial-mode degrees of freedom, each of which is randomly in one of the sixteen states $\{|H\rangle \otimes |b_1\rangle, |V\rangle \otimes |b_1\rangle, |R\rangle \otimes |b_1\rangle, |A\rangle \otimes |b_1\rangle, |H\rangle \otimes |b_2\rangle, |V\rangle \otimes |b_2\rangle, |R\rangle \otimes |b_2\rangle, |A\rangle \otimes |b_2\rangle, |H\rangle \otimes |s\rangle, |V\rangle \otimes |s\rangle, |R\rangle \otimes |s\rangle, |A\rangle \otimes |s\rangle, |H\rangle \otimes |a\rangle, |V\rangle \otimes |a\rangle, |R\rangle \otimes |a\rangle, |A\rangle \otimes |a\rangle\}$, and randomly inserts them into $S_T$ after her encryption to form sequence $S'_T$. Finally, Alice transmits $S'_T$ to Bob with the block transmission method.

After Bob confirms the receipt of $S'_T$, Alice and Bob implement the security check procedure similar to that of Step 1. If the transmission of $S'_T$ is secure, they will continue the next step; otherwise, they will halt the communication.

**Step 3: Bob's decryption and encoding.** Bob drops out the decoy single photons in $S'_T$ to restore $S_T$ after Alice's encryption. Afterward, Bob measures partciles $B_1^i$ and $B_2^i$ with $Z_P$ basis, respectively, where $i = 1, 2, \cdots, N$. Then, Bob decrypts the single-photon states in $S_T$ after Alice's encryption. That is, if the measurment results of partciles $B_1^i$ and $B_2^i$ are $|H\rangle$ and $|H\rangle$, respectively, Bob performs the composite unitary operation $C_{00}$ on the $i$ th single-photon states in $S_T$ after Alice's encryption; if the measurment results of partciles $B_1^i$ and $B_2^i$ are $|H\rangle$ and $|V\rangle$, respectively, Bob performs the composite unitary operation $C_{01}$ on the $i$ th single-photon states in $S_T$ after Alice's encryption; if the measurment results of partciles $B_1^i$ and $B_2^i$ are $|V\rangle$ and $|H\rangle$, respectively, Bob performs the composite unitary operation $C_{10}$ on the $i$ th single-photon states in $S_T$ after Alice's encryption; and if the measurment results of partciles $B_1^i$ and $B_2^i$ are $|V\rangle$ and $|V\rangle$, respectively, Bob performs the composite unitary operation $C_{11}$ on the $i$ th single-photon states in $S_T$ after Alice's encryption. Apparently, after Bob's decryption, $S_T$ is restored to $\{|\phi\rangle_1, |\phi\rangle_2, \cdots, |\phi\rangle_N\}$. Then, Bob measures $|\phi\rangle_i$ with the base $Z_P \otimes Z_S$ to know its initial state. According to his measurement result, Bob reproduces a new $|\phi\rangle_i$ with no measurement performed. In order to encode his two private classical bits $(k_i, l_i)$, Bob performs the composite unitary operation $C_{k_i l_i}$ on the new $|\phi\rangle_i$. Accordingly, the new $|\phi\rangle_i$ is turned into $C_{k_i l_i}|\phi\rangle_i$. As a result, $S''_T = \{C_{k_1 l_1}|\phi\rangle_1, C_{k_2 l_2}|\phi\rangle_2, \cdots, C_{k_N l_N}|\phi\rangle_N\}$ is formed. For the sake of security check, Bob prepares enough decoy single photons in both polarization and spatial-mode degrees of freedom, each of which is randomly in one of the sixteen states $\{|H\rangle \otimes |b_1\rangle, |V\rangle \otimes |b_1\rangle, |R\rangle \otimes |b_1\rangle, |A\rangle \otimes |b_1\rangle, |H\rangle \otimes |b_2\rangle, |V\rangle \otimes |b_2\rangle, |R\rangle \otimes |b_2\rangle, |A\rangle \otimes |b_2\rangle, |H\rangle \otimes |s\rangle, |V\rangle \otimes |s\rangle, |R\rangle \otimes |s\rangle, |A\rangle \otimes |s\rangle, |H\rangle \otimes |a\rangle, |V\rangle \otimes |a\rangle, |R\rangle \otimes |a\rangle, |A\rangle \otimes |a\rangle\}$, and randomly inserts them into $S''_T$ to form sequence $S'''_T$. Finally, Bob transmits $S'''_T$ to Alice with the block transmission method.

After Alice confirms the receipt of $S'''_T$, Alice and Bob implement the security check procedure similar to that of Step 1. If the transmission of $S'''_T$ is secure, they will continue the next step; otherwise, they will halt the communication.

**Step 4: Alice's encoding and bidirectional communication.** Alice drops out the decoy single photons in $S'''_T$ to restore $S''_T$.

Then, in order to encode her two private classical bits $(t_i, j_i)$, Alice performs the composite unitary operation $C_{t_i j_i}$ on $C_{k_i l_i}|\phi\rangle_i$. Accordingly, $C_{k_i l_i}|\phi\rangle_i$ is turned into $C_{t_i j_i} C_{k_i l_i}|\phi\rangle_i$. Afterward, Alice measures $C_{t_i j_i} C_{k_i l_i}|\phi\rangle_i$ with the base $Z_P \otimes Z_S$. For bidirectional communication, Alice publicly announces the measurement result of $C_{t_i j_i} C_{k_i l_i}|\phi\rangle_i$. According to the initial state of $|\phi\rangle_i$, her own composite unitary operation $C_{t_i j_i}$ and the measurement result of $C_{t_i j_i} C_{k_i l_i}|\phi\rangle_i$, Alice can read out Bob's two private classical bits $(k_i, l_i)$. Likewise, according to his knowledge about the initial state of $|\phi\rangle_i$, his own composite unitary operation $C_{k_i l_i}$ and Alice's announcement on the measurement result of $C_{t_i j_i} C_{k_i l_i}|\phi\rangle_i$, Bob can also read out Alice's two private classical bits $(t_i, j_i)$.

Now we use an example to further explain the bidirectional communication process. Suppose that $(t_1, j_1)$, $(k_1, l_1)$ and $|\phi\rangle_1$ are (0,0), (0,1) and $|H\rangle \otimes |b_1\rangle$, respectively. Alice prepares one quantum state $|\psi\rangle$ and transmits particles $B_1^1$ and $B_2^1$ to Bob. Then, Alice measures particles $A_1^1$ and $A_2^1$ with $Z_P$ basis and performs the coresponding composite unitary operation on $|\phi\rangle_1$ to encrypt it. Afterward, Alice transmits the encrypted $|\phi\rangle_1$ to Bob. Bob measures particles $B_1^1$ and $B_2^1$ with $Z_P$ basis and performs the coresponding composite unitary operation on the encrypted $|\phi\rangle_1$ to decrypt out $|\phi\rangle_1$. Then, Bob measures $|\phi\rangle_1$ with the base $Z_P \otimes Z_S$ to know its initial state. According to his measurement result, Bob reproduces a new $|\phi\rangle_1$ with no measurement performed. Hence, the new $|\phi\rangle_1$ evolves as follows:

$$|\phi\rangle_1 = |H\rangle \otimes |b_1\rangle \Rightarrow C_{01}|\phi\rangle_1 = (I_P \otimes U_S)(|H\rangle \otimes |b_1\rangle) = |H\rangle \otimes |b_2\rangle \Rightarrow C_{00} C_{01}|\phi\rangle_1 = (I_P \otimes I_S)(|H\rangle \otimes |b_2\rangle) = |H\rangle \otimes |b_2\rangle. \quad (13)$$

According to three pieces of known information: the announced measurement result $|H\rangle \otimes |b_2\rangle$, the prepared initial state $|H\rangle \otimes |b_1\rangle$ and her composite unitary operation $C_{00}$, Alice can read out that $(k_1, l_1)$ are (0,1). Similarly, Bob can read out that $(t_1, j_1)$ are (0,0), according to the announced measurement result $|H\rangle \otimes |b_2\rangle$, the prepared initial state $|H\rangle \otimes |b_1\rangle$ and his composite unitary operation $C_{01}$.

## 4 Security analysis
### (1) The active attacks from an outside attacker
We analyze the security against the active attacks from an outside attacker according to each step of the proposed QD protocol.

**Quantum key generation and distribution.** In this step, Alice prepares $N$ four-particle entangled states all in the state of $|\psi\rangle$ as the private quantum key, and sends $S'_{B_1}$ and $S'_{B_2}$ to Bob. Eve may try her best to obtain $S_{B_1}$ and $S_{B_2}$, and utilize them to decrypt the ciphertext Alice transmits to Bob subsequently. If Eve can succeed in doing it, she may know the initial states of $|\phi\rangle_i$. As a result, Eve may legally know the classical correlation [25] between Alice's two private classical bits $(t_i, j_i)$ and Bob's two private classical bits $(k_i, l_i)$ after Alice announces the measurement result of $C_{t_i j_i} C_{k_i l_i}|\phi\rangle_i$. Fortunately, Eve cannot acheive this goal. The reason lies in that Eve's attack on $S'_{B_1}$ or $S'_{B_2}$ will inevitably leave traces on decoy photons and be detected by Alice and Bob, since she cannot know the genuine positions of decoy photons. It is worthy of emphasizing that the decoy photon technology [39-40] can be considered as a variant of the security check approach of the BB84 QKD protocol [1], which has been validated to be unconditionally secure [41].

**Alice's encryption.** In this step, Alice encrypts $|\phi\rangle_i$ and transmits the ciphertext to Bob. If Eve wants to decrypt out the initial state of $|\phi\rangle_i$, she firstly needs to know the states of partciles $B_1^i$ and $B_2^i$ during the distribution process of quantum key. However, as analyzed above, the usage of decoy photon technology prevents Eve from acheiving this goal. In fact, Eve cannot intercept the ciphertext without being discovered, due to the decoy photon technology used in this step. Ref.[42] has verified that the decoy photon technology used in this step can effectively prevent the intercept-resend attack, the measure-resend attack and the entangle-measure attack from Eve. For example, according to Ref.[42], Eve's intercept-resend attack can be discovered by the security check process with the probability of $1 - \left(\frac{1}{4}\right)^{\delta_1}$, since Bob's measurement results on the fake decoy single photons are not always same to the genuine ones; Eve's measure-resend attack can be discovered by the security check process with the probability of $1-\left(\frac{9}{16}\right)^{\delta_1}$, since Eve's measuring bases for decoy single photons are not always identical to Alice's preparing bases. Here, $\delta_1$ is the number of decoy photons.

**Bob's decryption and encoding.** In this step, Bob decrypts out the initial state of $|\phi\rangle_i$, encodes his two private classical bits $(k_i, l_i)$ on it and transmits the particle $C_{k_i l_i}|\phi\rangle_i$ to Alice. Eve still cannot extract $(k_i, l_i)$ from $C_{k_i l_i}|\phi\rangle_i$ even if she intercepts it, as she has no access to the initial state of $|\phi\rangle_i$. In fact, Eve cannot intercept $C_{k_i l_i}|\phi\rangle_i$ without being detected, due to the usage of decoy photon technology in this step.

**Alice's encoding and bidirectional communication.** In this step, Alice encodes her two private classical bits $(t_i, j_i)$ on $C_{k_i l_i}|\phi\rangle_i$, and accomplishes the bidirectional communication process with Bob. Apparently, there are no photons transmitted in this step. As a result, Eve has no chance to launch her attack in this step.

**(2) The information leakage problem**

Without loss of generality, the example of Sect.3 is taken to analyze the information leakage problem here.

With the aid of private quantum key $|\psi\rangle$, Bob can definitely know the initial state of $|\phi\rangle_1$, thus it is unnecessary for Alice to publicly announce it. In this circumstance, Eve has no access to the initial state of $|\phi\rangle_1$. Hence, Eve has to guess it randomly when she hears the measurement result of $C_{00}C_{01}|\phi\rangle_1$ from Alice. If she guesses that the initial state of $|\phi\rangle_1$ is $|H\rangle \otimes |b_1\rangle$, she will think that $\{(t_1, j_1), (k_1, l_1)\}$ are one of $\{(0,0),(0,1)\}, \{(0,1),(0,0)\}, \{(1,0),(1,1)\}, \{(1,1),(1,0)\}$; if she guesses that the initial state of $|\phi\rangle_1$ is $|H\rangle \otimes |b_2\rangle$, she will think that $\{(t_1, j_1), (k_1, l_1)\}$ are one of $\{(0,0),(0,0)\}, \{(0,1),(0,1)\}, \{(1,0),(1,0)\}, \{(1,1),(1,1)\}$; if she guesses that the initial state of $|\phi\rangle_1$ is $|V\rangle \otimes |b_1\rangle$, she will think that $\{(t_1, j_1), (k_1, l_1)\}$ are one of $\{(0,0),(1,1)\}, \{(0,1),(1,0)\}, \{(1,0),(0,1)\}, \{(1,1),(0,0)\}$; and if she guesses that the initial state of $|\phi\rangle_1$ is $|V\rangle \otimes |b_2\rangle$, she will think that $\{(t_1, j_1), (k_1, l_1)\}$ are one of $\{(0,0),(1,0)\}, \{(0,1),(1,1)\}, \{(1,0),(0,0)\}, \{(1,1),(0,1)\}$. As a result, with respect to $\{(t_1, j_1), (k_1, l_1)\}$, there are sixteen kinds of probabilities for Eve, which contain $-\sum_{i=1}^{16} p_i \log_2 p_i = -\sum_{i=1}^{16} \frac{1}{16} \times \log_2 \frac{1}{16} = 4$ bit information, according to Shannon's information theory [43]. Thus, neither $(t_1, j_1)$ nor $(k_1, l_1)$ have been leaked out to Eve. It can be concluded that the proposed QD protocol can avoid the information leakage problem.

## 5 Discussions and conclusions
**(1) The information-theoretical efficiency**

The information-theoretical efficiency [3] is defined as $\gamma = \frac{b_s}{q_t + b_t} \times 100\%$, where $b_s$, $q_t$ and $b_t$ are the expected received private classical bits, the consumed qubits and the consumed classical bits for the communication between Alice and Bob. In the proposed QD protocol, without considering the security check procedures, $|\phi\rangle_i$ can be used to exchange Alice's two private classical bits $(t_i, j_i)$ and Bob's two private classical bits $(k_i, l_i)$ with two classical bits consumed for Alice's announcement on the measurement result of $C_{t_i j_i} C_{k_i l_i} |\phi\rangle_i$. In addition, one four-particle quantum entangled state $|\psi\rangle$ is used to encrypt $|\phi\rangle_i$, and Bob needs to reproduce a new $|\phi\rangle_i$ with no measurement performed. Therefore, the information-theoretical efficiency of the proposed QD protocol is $\gamma = \frac{2+2}{2+4+2+2} \times 100\% = 40\%$.

**(2) Comparisons of previous QD protocols with single photons in both polarization and spatial-mode degrees of freedom**

Recently, Wang et al. [36] put forward a QD protocol also with single photons in both polarization and spatial-mode degrees of freedom. Subsequently, Zhang and Situ [37] discovered that the QD protocol of Ref.[36] has the security loophole of information leakage, and suggested an improvement to avoid this problem through modifying Alice and Bob's encoding rules. In Ref.[37], without considering the security check procedures, one single photon in both polarization and spatial-mode degrees of freedom can be utilized to mutually transmit one classical bit from Alice and one classical bit from Bob, while two classical bits are needed for Alice's announcement on the value of $R$. Hence, the information-theoretical efficiency of Ref.[37] is $\gamma = \frac{2}{2+2} \times 100\% = 50\%$. However, in Ref.[37], one single photon in both polarization and spatial-mode degrees of freedom can only totally carry two classical bits, while it can totally carry four classical bits in the proposed QD protocol. Hence, although the information-theoretical efficiency of the proposed QD protocol is a little smaller than that of Ref.[37], its quantum communication capacity is twice that of Ref.[37].

In addition, we also proposed an information leakage resistant QD protocol with single photons in both polarization and spatial-mode degrees of freedom in Ref.[42] recently. In Ref.[42], without considering the security check procedures, two adjacent single photons can be utilized to exchange two classical bits from Alice and two classical bits from Bob, while four classical bits are needed for Bob's announcement on his measurement result of the final state of encoded single photon. Hence, the information-theoretical efficiency of Ref.[42] is $\gamma = \frac{4}{4+4} \times 100\% = 50\%$. However, Ref.[42] needs to generate two adjacent single photons in both polarization and spatial-mode degrees of freedom in the same state, which may be hard to achieve in practical circumstance. Although the information-theoretical efficiency of the proposed QD protocol is a little smaller than that of Ref.[42], it doesn't have the trouble of generating two adjacent single photons in both polarization and spatial-mode degrees of freedom in the same state.

In conclusion, in this paper, we put forward a novel information leakage resistant QD protocol with single photons in both polarization and spatial-mode degrees of freedom, which utilizes quantum encryption technology to overcome the information leakage problem. In the proposed QD protocol, during the transmission process, the single photons in both polarization and spatial-mode degrees of freedom used for encoding two communicants' private classical bits are protected by both quantum encryption technology and decoy photon technology. The initial states of the single photons in both polarization and spatial-mode degrees of freedom used for encoding two communicants' private classical bits are shared between two communicants through quantum key encryption and decryption. As a result, the information leakage problem is avoided. The information-theoretical efficiency of the proposed QD protocol is as high as 40%. The proposed QD protocol has twice quantum communication capacity of Zhang and Situ's QD protoocl in Ref.[37], and needn't to prepare two adjacent single photons in both polarization and spatial-mode degrees of freedom in the same state. In the future, for further enhancing the information-theoretical efficiency, we will be devoted to researching how to make the quantum key repeatedly used.

### Acknowledgements

Funding by the National Natural Science Foundation of China (Grant No.62071430) and Zhejiang Gongshang University, Zhejiang Provincial Key Laboratory of New Network Standards and Technologies (No. 2013E10012) is gratefully acknowledged.

### References

[1] Bennett C H, Brassard G. Quantum cryptography: public-key distribution and coin tossing. In: Proceedings of the IEEE International Conference on Computers, Systems and Signal Processing. Bangalore: IEEE Press, 1984, 175-179
[2] Bennett C H, Brassard G, Mermin N D. Quantum cryptography without Bell theorem. Phys Rev Lett, 1992, 68:557-559
[3] Cabello A. Quantum key distribution in the Holevo limit. Phys Rev Lett, 2000, 85:5635
[4] Beige A, Englert B G, Kurtsiefer C,et al.. Secure communication with a publicly known key. Acta Phys Pol A,2002,101: 357-368
[5] Yang Y G, Wen Q Y. An efficient two-party quantum private comparison protocol with decoy photons and two-photon entanglement. J Phys A: Math and Theor, 2009, 42(5): 055305
[6] Yang Y G, Gao W F, Wen Q Y. Secure quantum private comparison. Phys Scr, 2009, 80(6): 065002
[7] Yang Y G, Xia J, Jia X, Shi L, Zhang H. New quantum private comparison protocol without entanglement. Int J Quant Inform, 2012, 10(6):1250065
[8] Liu B, Gao F, Jia H Y, Huang W, Zhang W W, Wen Q Y. Efficient quantum private comparison employing single photons and collective detection. Quantum Inf Process, 2013,12(2):887-897
[9] Ye T Y. Quantum private comparison via cavity QED. Commun Theor Phys, 2017, 67 (2) :147-156
[10] Long G L, Liu X S. Theoretically efficient high-capacity quantum-key-distribution scheme. Phys Rev A, 2002,65: 032302
[11] Bostrom K, Felbinger T. Deterministic secure direct communication using entanglement. Phys Rev Lett, 2002, 89:187902
[12] Deng F G, Long G L, Liu X S.Two-step quantum direct communication protocol using the Einstein-Podolsky-Rosen pair block. Phys Rev A, 2003, 68:042317
[13] Deng F G, Long G L. Secure direct communication with a quantum one-time pad. Phys Rev A, 2004, 69: 052319
[14] Wang C, Deng F G, Li Y S, Liu X S, Long G L. Quantum secure direct communication with high-dimension quantum superdense coding. Phys Rev A, 2005,71:044305
[15] Zhang Z J, Man Z X. Secure direct bidirectional communication protocol using the Einstein-Podolsky-Rosen pair block. 2004,http://arxiv.org/pdf/quant-ph/0403215.pdf
[16] Zhang Z J,Man Z X. Secure bidirectional quantum communication protocol without quantum channel. 2004, http://arxiv.org/pdf/quant-ph/0403217.pdf
[17] Nguyen B A. Quantum dialogue. Phys Lett A,2004, 328(1):6-10
[18] Jin X R, Ji X, Zhang Y Q,Zhang S,et al..Three-party quantum secure direct communication based on GHZ states. Phys Lett A, 2006,

354(1-2): 67-70
[19] Man Z X, Xia Y J. Controlled bidirectional quantum direct communication by using a GHZ state. Chin Phys Lett,2006, 23(7): 1680-1682
[20] Ji X, Zhang S. Secure quantum dialogue based on single-photon. Chin Phys, 2006, 15(7):1418-1420
[21] Man Z X, Xia Y J, Nguyen B A. Quantum secure direct communication by using GHZ states and entanglement swapping. J Phys B-At Mol Opt Phys, 2006, 39(18):3855-3863
[22] Yang Y G, Wen Q Y. Quasi-secure quantum dialogue using single photons. Sci China Ser G- Phys Mech Astron, 2007,50(5):558-562
[23] Shan C J, Liu J B, Cheng W W, Liu T K, Huang Y X, Li H. Bidirectional quantum secure direct communication in driven cavity QED. Mod Phys Lett B, 2009,23(27):3225-3234
[24] Ye T Y, Jiang L Z. Improvement of controlled bidirectional quantum secure direct communication by using a GHZ state. Chin Phys Lett, 2013, 30(4):040305
[25] Tan Y G, Cai Q Y. Classical correlation in quantum dialogue. Int J Quant Inform, 2008, 6(2):325-329
[26] Gao F, Qin S J, Wen Q Y,Zhu F C. Comment on: "Three-party quantum secure direct communication based on GHZ states". Phys Lett A, 2008,372(18):3333-3336
[27] Gao F,Guo F Z,Wen Q Y,Zhu F C.Revisiting the security of quantum dialogue and bidirectional quantum secure direct communication. Sci China Ser G- Phys Mech Astron, 2008,51(5):559-566
[28] Shi G F, Xi X Q, Tian X L, Yue R H. Bidirectional quantum secure communication based on a shared private Bell state. Opt Commun, 2009,282(12):2460-2463
[29] Shi G F, Xi X Q, Hu M L, Yue R H. Quantum secure dialogue by using single photons. Opt Commun, 2010, 283(9): 1984-1986
[30] Shi G F. Bidirectional quantum secure communication scheme based on Bell states and auxiliary particles. Opt Commun, 2010, 283(24):5275-5278
[31] Gao G. Two quantum dialogue protocols without information leakage. Opt Commun, 2010, 283(10):2288-2293
[32] Ye T Y. Large payload bidirectional quantum secure direct communication without information leakage. Int J Quant Inform, 2013, 11(5): 1350051
[33] Ye T Y. Quantum secure dialogue with quantum encryption. Commun Theor Phys, 2014, 62(3):338-342
[34] Huang L Y, Ye T Y. A kind of quantum dialogue protocols without information leakage assisted by auxiliary quantum operation. Int J Theor Phys, 2015, 54(8):2494-2504
[35] Liu D, Chen J L, Jiang W. High-capacity quantum secure direct communication with single photons in both polarization and spatial-mode degrees of freedom. Int J Theor Phys,2012,51:2923-2929
[36] Wang L L, Ma W P, Shen D S, Wang M L. Efficient bidirectional quantum secure direct communication with single photons in both polarization and spatial-mode degrees of freedom. Int J Theor Phys, 2015, 54:3443-3453
[37] Zhang C, Situ H Z. Information leakage in efficient bidirectional quantum secure direct communication with single photons in both polarization and spatial-mode degrees of freedom. Int J Theor Phys, 2016, 55:4702-4708
[38] Wang L L, Ma W P, Wang M L, Shen D S. Three-party quantum secure direct communication with single photons in both polarization and spatial-mode degrees of freedom. Int J Theor Phys, 2016, 55:2490-2499
[39] Li C Y, Zhou H Y,Wang Y,Deng F G. Secure quantum key distribution network with Bell states and local unitary operations. Chin Phys Lett, 2005,22(5):1049-1052
[40] Li C Y,Li X H,Deng F G,Zhou P,Liang Y J,Zhou H Y. Efficient quantum cryptography network without entanglement and quantum memory. Chin Phys Lett, 2006, 23(11):2896-2899
[41] Shor P W, Preskill J. Simple proof of security of the BB84 quantum key distribution protocol. Phys Rev Lett, 2000, 85(2):441
[42] Ye T Y, Li H K, Hu J L. Information leakage resistant quantum dialogue with single photons in both polarization and spatial-mode degrees of freedom. Quantum Inf Process, 2021, 20(6): 209
[43] Shannon C E. Communication theory of secrecy system. Bell System Tech J,1949,28:656-715